\documentclass[prc,12pt,aps,showpacs,preprint,
tightenlines,
superscriptaddress]{revtex4}
\usepackage{epsf}
\usepackage{amsmath}

\def\beq{\begin{equation}}
\def\eeq{\end{equation}}
\def\bea{\begin{eqnarray}}
\def\eea{\end{eqnarray}} 
\def\beqa{\begin{equation}\begin{array}{l}}
\def\eeqa{\end{array}\end{equation}}

\def\eqlab#1{\label{eq:#1}}
\def\figlab#1{\label{fig:#1}}
\def\tablab#1{\label{tab:#1}}

\def\eref#1{(\ref{eq:#1})}
\def\eqref#1{eq.~(\ref{eq:#1})}
\def\Eqref#1{Eq.~(\ref{eq:#1})}

\def\Figref#1{Fig.~\ref{fig:#1}}
\def\tabref#1{\ref{tab:#1}}

\def\sla#1{#1 \!\!\!\! \slash}
\def\slad{\partial \!\!\! \slash}
\def\slap{p \!\!\! \slash}
\def\sleps{\veps \!\!\!\slash}

\def\half{\mbox{\small{$\frac{1}{2}$}}}
\def\thalf{\mbox{\small{$\frac{3}{2}$}}}

\def\third{\mbox{\small{$\frac{1}{3}$}}}

\def\barr{\left(\begin{array}{c}}
\def\earr{\end{array}\right)}
\def\bmat{\left(\begin{array}{cc}}
\def\emat{\end{array}\right)}
\def\al{\alpha}
\def\be{\beta}
\def\ga{\gamma} \def\Ga{{\it\Gamma}}
\def\de{\delta} \def\De{\Delta}
\def\veps{\varepsilon}  

\def\la{\lambda} \def\La{{\it\Lambda}}

\def\Psit{{\it\Psi}}

\def\w{\omega}

\def\pa{\partial}

\def\ie{{i.e., }}

\def\pa{\partial}
\def\ra{\rightarrow}
\def\no{\nonumber}
\def\CF#1#2#3#4{#1 {\bf #2}, #4 (#3)}  
\def\ibid {{\it ibid.}}
\def\etal {{\it et~al.}}
\def\ann {Ann.~Phys.~(NY)}

\def\prc {Phys. Rev.~C}

\def\sjnp {Sov.~J.~Nucl.~Phys.}

\def\rg{{\rm g}} \def\rp{{\rm p}}

 \def\rn{{\rm n}}

\def\lag{{\mathcal L}}

\def\MA{{\mathcal A}}
\def\MB{{\mathcal B}}

\def\mathscr{\mathcal}

\def\bGa{{\bf\Gamma}}

 \def\Psib{\bar{\Psit}}

\def\vp{\vec{p}}

\def\pn{$\pi N$ }

\def\im{\mbox{Im}}
\def\3d{3-D}

\def\ol#1{\overline{#1}} 

\begin{document}
\preprint{JLAB-THY-04-249, WM-04-109}
\title{Pion photoproduction on nucleons in a covariant hadron-exchange model}

\author{Vladimir Pascalutsa}
\email{vlad@jlab.org}

\affiliation{Theory Group, Jefferson Laboratory, 12000 Jefferson Ave, Newport News, VA 23606}
\affiliation{Department of Physics, The College of William \& Mary, Williamsburg, VA 23188}
\affiliation{Department of Physics and Astronomy, Ohio
University, Athens, OH 45701}
\author{John~A.~Tjon}
\email{tjon@jlab.org}
\affiliation{Theory Group, Jefferson Laboratory, 12000 Jefferson Ave, Newport News, VA 23606}
\affiliation{Department of Physics, University of Maryland, College Park, MD 20742}
%
%
\date{\today}

\thispagestyle{empty}

\begin{abstract}
We present a relativistic dynamical model of pion photoproduction on the nucleon
in the resonance region. It offers several advances over the existing approaches.
The model is obtained by extending our $\pi N$-scattering description
to the electromagnetic channels. The resulting photopion amplitude
is thus unitary in the $\pi N$, $\ga N$ channel space, Watson's theorem is exactly 
satisfied.
At this stage we have included the pion, nucleon, $\De(1232)$-resonance
degrees of freedom. The $\rho$ and $\omega$ meson exchanges are also included,
but play a minor role in the considered energy domain (up to $\sqrt{s}=1.5$ GeV) .
In this energy range the model provides a good description of all the important multipoles. 
We have allowed for only two free parameters --- the photocouplings of the $\Delta$-resonance.
These couplings  are adjusted to reproduce the
strength of corresponding resonant-multipoles $M_{1+}$ and $E_{1+}$ at the
 resonance position.  
\end{abstract}

\maketitle

\section{Introduction}
In recent years there has been significant interest in the pion and kaon {\it photo-} and 
{\it electroproduction}
off the nucleon. Several  excellent experimental programs exploring these reactions 
in the {\it resonance region} have been performed at MAMI, MIT Bates, BNL, and Jefferson Lab. 
To extract the resonance properties from the  experimental data a number of sophisticated tools
have been developed over the past decade. Most widely exploited are the {\it partial-wave}
photoproduction {\it solutions} SAID~\cite{SAID} and MAID~\cite{MAID}, {\it K-matrix
models}~\cite{Kmat},
and  {\it dynamical models}, such
as DMT~\cite{DMT}, the model of Sato and Lee~\cite{Sato96}, Gross and Surya model~\cite{GrS96},
and a number of others~\cite{others}.

In this paper we present a new dynamical model for {\it pion photoproduction}.
It is an extension of our model of pion-nucleon ($\pi$N) interaction~\cite{PaT98,PaT00} to include the
electromagnetic interaction in a way consistent with the Watson theorem~\cite{Wat54} and
current conservation. The framework is based on solving a Bethe-Salpeter type
of equation for the scattering amplitude in the channel space spanned by
$\pi N$ and $\ga N$ states. As in the $\pi$N case we use the {\it equal-time} (instanteneous)
quasipotential reduction of the Bethe-Salpeter equation. The driving force of the equation to lowest
order in interactions is given by single-particle exchange graphs. Here we
will restrict our discussion to the force given by the single  nucleon, pion, $\rho$- and $\w$-meson,
and $\De$(1232)-resonance exchanges. 

In comparing with other approaches based on the hadron-exchange dynamics
we note that they differ mainly in the use of  relativistic dynamics and the renormalisation
procedures. 
Our model bares close analogies to the relativistic model developed by
Gross and Surya~\cite{GrS96}. 
In contrast to their work, we
do not approximate the hadron exchanges in the $t$- and $u$-channel by
a separable interaction. As a result 
the integral equation for the $\pi N$ amplitude can be solved only numerically,
and hence the task of solving the model is more technically involved. 
Our models are also different in the choice of quasipotential reduction
of the Bethe-Salpeter equation --- equal-time {\it vs.} pion spectator. 

There are important differences of our model with the DMT model. First of all,
the dressed resonance-exchanges in the $s$-channel are represented by a Breit-Wigner form with
``unitarity phases'' which need to be fitted to the condition of Watson's theorem.
In our model these contributions are generated dynamically. This has an advantage
of satisfying Watson's theorem automatically. Also, the resonance parameters, apart from
the electromagnetic couplings, are thus fully constrained by the $\pi$N interaction and need not
to be fitted separately. Second major difference is again in the choice of relativistic dynamics
--- DMT model exploits the Kadyshevsky quasipotential reduction of the Bethe-Salpeter equation.

Sato and Lee~\cite{Sato96} apply the Hamiltonian approach and the method of unitary transformations which
makes it difficult to compare directly to the Bethe-Salpeter type of approach. The generic feature
that distinguishes the two is that in the quantum-mechanical Hamiltonian description the
particles are always on the mass shell and intermediate particles are off the energy shell, while
in the field-theoretic desription it is the other way around. Another difference is that Sato and Lee
do not perform any renormalizations of the dressed baryon-pole contributions.

In this paper we shall only briefly present the framework and the results
for the photoproduction multipoles.  The results for the pion-photoproduction observables as well
as the extension to electroproduction of pions will appear in
subsequent publications, see e.g.~\cite{CPTW04}. The paper is organized as follows.
In the following section we briefly summarize the usual arguments for inclusion 
of the $\pi N$ final state interaction in the $\pi$ photoproduction reactions. 
In Sect.~III we construct the pion-photoproduction potential with emphasis on satisfiyng
the gauge-invariance constraints.
In Sect.~IV the renormalization of the pole
terms of the photoproduction amplitude is described. 
In Sect.~V we present some numerical results for 
the pion-photoproduction multipoles and discussion. Sect.~VI concludes the paper.

\section{$\pi N$-$\ga N$ coupled channel equations}

To include the photon 
in a way preserving unitarity in the channel space
spanned by  the $\pi N$ and $\ga N$ states we consider the following four processes:
\beq
\begin{array}{lr}
\pi N\ra \pi N ,\,\,  & \pi N \ra\ga N, \,\,\\
\ga N \ra \pi N, \,\,& \ga N \ra \ga N \,.
\end{array}  
\eeq
and the following  coupled-channel
scattering equation,
\beq
\eqlab{channel}
\bmat T_{\pi\pi} & T_{\pi \ga}\\
T_{\ga \pi} & T_{\ga\ga} \emat
=\bmat V_{\pi\pi} & V_{\pi \ga}\\
V_{\ga \pi} & V_{\ga\ga} \emat +
 \bmat V_{\pi\pi} & V_{\pi \ga}\\
V_{\ga \pi} & V_{\ga\ga} \emat
\bmat G_{\pi} & 0\\
0& G_{\ga} \emat
\bmat T_{\pi\pi} & T_{\pi \ga}\\
T_{\ga \pi} & T_{\ga\ga} \emat,
\eeq 
where $T$ and $V$ are the suitably normalized amplitudes and 
driving potentials of the \pn scattering ($\pi\pi$), 
pion photo-production ($\pi\ga$),
absorption ($\ga\pi$), and the nucleon Compton effect ($\ga\ga$). 
The propagators $G_\pi$ and $G_\ga$ are, respectively, 
the pion-nucleon and photon-nucleon two-particle propagators.
With the assumption of hermiticity of the potential and time-reversal
symmetry which in particular relates the $\ga\pi$ and $\pi\ga$
amplitudes, \Eqref{channel} leads to an exactly unitary
$S$-matrix, $S_{fi}=\de_{fi}+2i T_{fi}$, in the defined channel space.

Since iterations of the potentials involving the photon give rise to the 
small electromagnetic corrections, one can simplify
the equation by keeping only  the leading order in the electric charge
$e$. This leads to 
\begin{subequations}
\bea
\eqlab{trunc}
\eqlab{eq1}
T_{\pi\pi} &=& V_{\pi\pi} + V_{\pi\pi} G_{\pi} T_{\pi\pi},\\
\eqlab{eq2}
T_{\pi\ga} &=& V_{\pi\ga} + T_{\pi\pi} G_{\pi} V_{\pi\ga},\\
T_{\ga\pi} &=& V_{\ga\pi} + V_{\ga\pi} G_{\pi} T_{\pi\pi},\\
T_{\ga\ga} &=& V_{\ga\ga} + V_{\ga\pi} G_{\pi} T_{\pi\ga}. 
\eea
\end{subequations}
In this approximation, the integral equation has to be
solved for the \pn amplitude only. The rest is determined in a
one-loop calculation. 

The equation for the \pn amplitude, \Eqref{eq1}, has been studied by us previously in the framework
of relativistic quasipotential scattering with the \pn potential modeled
by a number of relevant hadron exchanges~\cite{PaT00}.  The parameters have been fitted to the \pn-scattering
partial-wave analysis data. 
In the present work we obtain the photoproduction amplitude from \Eqref{eq2} 
(diagrammatically shown in \Figref{gipif}) using exactly the same 
quasipotential approach and the \pn amplitude as in Ref.~\cite{PaT00}.
 The only free parameters in this calculation will
be the electromagnetic couplings of hadrons entering the driving force $V_{\pi\ga}$,
all the rest is fixed through the analysis of \pn scattering.

Our resulting photoproduction amplitude obeys
the {\it Watson theorem}~\cite{Wat54},
which relates the phase of the photoproduction amplitude to the
\pn elastic phase shift $\de_{\pi\pi}$:
\beq
T_{\ga\pi} = |T_{\ga\pi}|\, e^{i\de_{\pi\pi}}.
\eeq
The phase of the photoproduction amplitude is thus
fully determined in terms of the on-shell $\pi N$ amplitude.
The dependence on the off-shell behavior of the $\pi N$ interaction
resides fully in the absolute magnitude of the photoproduction
amplitude.

\section{The model potential and gauge invariance}

The pion-photoproduction potential of this model is shown diagrammatically
in \Figref{gapipotf}. The first four graphs represent the
so-called Born term, where  the fourth graph 
is the Kroll-Ruderman contact term. The latter is
obtained by the ``minimal substitution'' in the pseudo-vector $\pi NN$
coupling, and is therefore needed to ensure the current conservation
of the Born contribution.

Except for the
$\ga N\De$ couplings, the Lagrangian we use is standard. 
For brevity we only specify here the Feynman rules for corresponding 
vertices [omitting isospin, the isospin structure will be specified below, c.f.~\Eqref{isosp}]:
\begin{subequations}
\bea
&& \Ga_{\ga NN}^\mu (\kappa;q) = e\ga^\mu 
- \frac{e\kappa}{2m_N}\ga^{\mu\nu}q_\nu, \\
&&\Ga_{\ga \pi NN}^\mu = \frac{e g_{\pi NN}}{2m_N}\ga^\mu\ga_5, \\
&&\Ga_{\ga\pi\pi}^\mu(k',k) = e (k'+k)^\mu, \\
&&  \Ga_{\ga\pi v}^{\mu\al} (q,k) = \frac{e g_{\ga\pi v}}{m_\pi}
\veps^{\mu\al\be\nu} k_\be q_\nu,
\eea
\end{subequations}
where $e$ is the proton electric charge
($e^2/4\pi\simeq 1/137 $),
$\kappa $ is the anomalous magnetic moment of the nucleon, $m_N\simeq 0.9383$ GeV and
$m_\pi \simeq 0.139$ GeV are
the nucleon and pion masses, $\ga^{\mu\nu}=\half [\ga^\mu,\ga^\nu]$,
$q$ and $k$ denote the momenta of the photon and pion, respectively.
The subscript $v$ stands for a vector meson, in this case $\rho$ or $\w$. 

The $\ga N\De$ vertices we obtain from the following Lagrangian,
\beq
\eqlab{ganDe}
\lag_{\ga N \De}= \frac{3\,e}{2 m_N (m_N+m_\De)}\,\ol N\, T_3^\dagger
\left(i g_M  \tilde F^{\mu\nu}
- g_E \gamma_5 F^{\mu\nu}\right)\,\pa_{\mu}\De_\nu
+ \mbox{H.c.},
\eeq
where $m_\De\simeq 1232$ MeV is the $\De$-isobar mass, $T_3$ is the isospin $N\De$ transition
matrix, with normalization $T^\dagger_3 T_3=\frac{2}{3}$.
This $\ga N\De$ coupling is  invariant under electromagnetic gauge 
transformations (to the order to which we work), as well as
under  the spin-3/2 gauge transformation:
\begin{equation}
\De_\mu(x) \rightarrow \De_\mu(x) + \pa_\mu \veps(x),
\label{eq:spin3/2gt}
\end{equation}
with $\veps$ a spinor field. Invariance under (\ref{eq:spin3/2gt}) ensures
the correct spin-degrees-of-freedom counting~\cite{Pa98}.  
In the $\Delta$'s rest frame  (where $\De_0=0$, $ \pa_0 \De_i=-i m_\De \De_i$,  
and $\pa_i \De_j=0$), the coupling \eref{ganDe} becomes
\beq
\eqlab{ganDe2}
\lag_{\ga N \De}=  - \frac{3\,e m_\De}{2 m_N( m_N+m_\De)}\,\ol N\, T_3^\dagger
\left( g_M  B^i + g_E \gamma_5 E^i \right)\, \De_i
+ \mbox{H.c.},
\eeq
where $B^i$ is the magnetic and $E^i$ the electric field.
Therefore the two terms correspond to  $N\De$ magnetic
and electric transitions, respectively. 

However, in the standard convention~\cite{JoS73}, the electric coupling $G_E$
is defined as the one directly proportional to the
$E_{1+}^{(3/2)}$ multipole. On the mass shell of the $\De$, our convention
and the convention of Jones and Scadron~\cite{JoS73} are related as follows:
\begin{subequations}
\eqlab{ff_Vlad}
\begin{eqnarray}
\eqlab{G1_Vlad} g_{M}&=&G_{M}-G_{E},\\
\eqlab{G2_Vlad}
g_E&=&-2\,G_E \,\frac{m_{\Delta}+m_N}{m_{\Delta}-m_N}  .
\end{eqnarray}
\end{subequations}

The Feynman rule corresponding to the coupling~\eref{spin3/2gt} reads:
\beq
\Ga_{\ga N\De}^{\al\mu} (p,q) = \frac{3 e}{2m_N(m_N+m_\De)}\left[ g_M\, 
\veps^{\al\mu\be\nu} p_\be q_\nu 
- g_E \, (p\cdot q \, g^{\al\mu} - q^\al p^\mu)\, i\ga_5\right] ,
\eeq
with  $p$ ($q$)  being the 4-momentum of  the $\De$ (photon),
and $\al$ ($\mu$) the vector index of the $\De$ (photon) field.

For the ``strong-interaction'' vertices we use the same forms as in \cite{PaT00}, namely:
\begin{subequations}
\bea
&& \Ga_{\pi NN}(k) = \frac{g_{\pi NN}}{2 m_N} \ga_5 \sla{k} ,\\
&& \Ga_{\pi NN^\ast}(k) = \frac{g_{\pi NN^\ast}}{2 m_N} \ga_5 \sla{k} ,\\
&& \Ga_{v NN}^\al (q) = g_{vNN} \left( \ga^\al 
- \frac{\kappa_v}{2m_N}\ga^{\al\nu}q_\nu \right), \\
&& \Ga_{\pi N\De}^\al(k,p) = \frac{f_{\pi N\De}}{m_\pi m_\De}
\veps^{\al\be\mu\nu} p_\be \ga_\mu \ga_5 k_\nu . 
\eea
\end{subequations}

The Feynman graphs depicted in \Figref{gapipotf} correspond to
\begin{subequations}
\bea
(4\pi)V^{({\rm S,V})\mu}_{\rm(N),pole} &=&
\Ga_{\pi NN}(k')\, S_N(p+q)\,\Ga_{\ga NN}^\mu (\kappa_{{\rm S,V}};q) \\
(4\pi)V^{({\rm S,V})\mu}_{\rm(N),exch} &=& \Ga_{\ga NN}^\mu (\kappa_{{\rm S,V}};q)\, 
S_N(p-k')\,\Ga_{\pi NN}(k') \\ 
(4\pi)V^{\mu}_{\rm(\De),pole} &=&
\Ga_{\pi N\De}^\al (k',p)\, S_\De^{\al\be}(p+q)
\,\Ga_{\ga N\De}^{\mu\be} (p;q) \\
(4\pi)V^{\mu}_{\rm(\De),exch} &=& \Ga_{\ga N\De}^{\mu\al} (p;q)
\, S_\De^{\al\be}(p-k')\, \Ga_{\pi N\De}^\be (k',p)\\
(4\pi)V^{\mu}_{\rm(\pi)} &=& \Ga_{\pi NN} (q-k')\, 
S_\pi(q-k')\,\Ga_{\ga\pi\pi}^\mu (k',q-k') \\ 
(4\pi)V^{\mu}_{\rm(KR)} &=& \Ga_{\ga\pi NN}^\mu \\
(4\pi)V^{\mu}_{(v)} &=& \Ga_{v NN}^\al (q-k')\, 
S_v^{\al\be}(q-k')\,\Ga_{\ga\pi v}^{\mu\be} (q,k'),\,\,\, v=(\rho,\w). 
\eea
\end{subequations}

These graphs give
the following contributions to the isospin $\pi \ga$ amplitudes:
\begin{subequations}
\eqlab{isosp}
\bea
V^{(1/2)\mu} &=& 3 V^{\rm(V)\mu}_{\rm(N),pole} - V^{\rm(V)\mu}_{\rm(N),exch}
+ 2 V^{\mu}_{(\pi)} + 2 V^{\mu}_{\rm(KR)} 
+\mbox{$\frac{4}{3}$} V^{\mu}_{\rm(\De),exch} + V^{\mu}_{(\w)}, \\
V^{(3/2)\mu} &=& 2 V^{\rm(V)\mu}_{\rm(N),exch}
- V^{\mu}_{(\pi)} - V^{\mu}_{\rm(KR)}
+ V^{\mu}_{\rm(\De),pole}
+ \half V^{\mu}_{\rm(\De),exch} + V^{\mu}_{(\w)}, \\
V^{(0)\mu} &=& V^{\rm(S)\mu}_{\rm(N),pole} + V^{\rm(S)\mu}_{\rm(N),exch}
+V^{\mu}_{(\rho)},
\eea 
\end{subequations}

The gauge invariance of the electromagnetic interactions imposes the following
{\it current conservation} condition,
\beq
q_\mu V^{(I)\mu} =0,
\eeq 
for all values of the isospin: $I=0,\half,\thalf$. 
For the $\De,\,\rho$
and $\w$ exchange graphs this condition is trivially satisfied, since
the corresponding electromagnetic vertices vanish when contracted with
the photon momentum. For the nucleon and pion exchange contributions
the situation is complicated by the fact the the photon couples
minimally and hence the vertices fulfill the Ward-Takahashi (WT) identities:
\begin{subequations}
\bea
&& \mbox{$\frac{1}{e}$} q_\mu \Ga_{\ga NN}^\mu (p',p)=
S_N^{-1} (p') - S_N^{-1} (p),\\
&& \mbox{$\frac{1}{e}$} q_\mu \Ga_{\ga\pi\pi}^\mu (k',k)=S_\pi^{-1} (k')
 - S_\pi^{-1} (k).
\eea    
\end{subequations}
Nonetheless, using these identities, it is easy to see that 
a cancellation among the nucleon, pion and the KR contact term
contributions leads to 
\bea
\eqlab{netcc}
(4\pi)\, q_\mu  
V^{(1/2)_\mu} &=& 
- 3 e \Ga_{\pi NN}(k')\, S_N(p+q)\,S_N^{-1} (p) 
+ e S_N^{-1} (p')\,S_N(p-k')\,\Ga_{\pi NN}(k')\no\\
&+& 2 e \Ga_{\pi NN} (q-k')\, 
S_\pi(q-k')\,S_\pi^{-1}(k'),
\eea
and analogously for the other isospin amplitudes.
Therefore, the current is conserved up to the terms proportional
to the inverse propagators of the external particles, and hence is
exactly conserved when the external particles are on the mass shell.
  
A problem arises when we would like to include the 
{\it sideways} form factors, i.e., cutoff functions dependent on the
4-momentum of the exchanged particle.  
Obviously, simply introducing them into the pion
and nucleon exchange graphs, as we have done it for the \pn potential in~\cite{PaT00},
will destroy the current conservation. The cancellation among the graphs does not anymore 
take place.

The easiest way to implement such cutoff form factors without loss of
current conservation is to perform
the minimal substitution on the form factors themselves. 
We follow essentially the method of Gross and Riska~\cite{GrR87}.
We use the fact that our sideways form factors depend exclusively
on the momentum of the exchanged particle and hence it makes no difference whether to
include the form factor into the vertex function or the inverse
form factor squared into the propagator. In the latter case
the minimal substitution is more straightforward. 

More specifically, in the nucleon case we start with
\beq
\lag = [f^{-1} (\pa^2)\Psib] (i\slad - m_N) f^{-1} (\pa^2)\Psit,
\eeq 
where $f(\pa^2)$ is the form factor operator in the coordinate space.
Substituting  $\pa_\mu$ by $D_\mu=\pa_\mu-ieA_\mu$, and linearizing
in the electromagnetic field we find the {\it modified} $\ga NN$ vertex:
\bea
 \Ga_{\ga NN}^{\mu,\rm{mod}}(p',p) &=& e\ga^\mu 
f^{-1}({p'}^2) f^{-1}(p^2)  + e\,(p+p')^\mu\no\\
&\times& [f^{-1}({p'}^2) (\slap'-m_N)
+ f^{-1}(p^2)(\slap-m_N)]\, \Xi({p'}^2,p^2)
\eea
where in general 
$\Xi$ is the finite difference  the inverse form factor:
\beq
\eqlab{xidef}
\Xi({p'}^2,p^2)=\frac{f^{-1}({p'}^2)-f^{-1}(p^2)}{{p'}^2-p^2}.
\eeq
For instance, for the monopole type [i.e, $f^{-1}(p^2)=(\La^2-p^2)/(\La^2-m^2)$]
we have simply $\Xi=(\La^2-m^2)^{-1}$. 

Taking the specific nucleon form factor used in
our \pn model,
\beq
\eqlab{formfnn}
f(p^2)=\left( \frac{2 \La_N^4}{2\La_N^4+(p^2-m_N^2)^2} \right)^2
\eeq 
we find
$\Xi({p'}^2,p^2)= ({p'}^2+p^2-2m)\,[f^{-1}({p'}^2)+f^{-1}(p^2)]/(2\La_N^4)$.

The anomalous magnetic moment term, $\Ga^\mu_{\rm amm}=
-(e\kappa/4m_N) [\ga^{\mu},\ga^{\nu}] q_\nu $, is explicitly gauge-invariant
and we choose to leave it unchanged. 
Adding it to the vertex, and substituting \Eqref{xidef} for 
$\Xi$, we obtain
\bea
\eqlab{modvert}
 \Ga_{\ga NN}^{\mu,\rm{mod}}(p',p) &=& e
f^{-1}({p'}^2) f^{-1}(p^2)
\left(\rg^{\mu\nu}- \frac{(p+p')^\mu q^{\nu}}{q\cdot(p+p')}\right)
\ga_\nu\no\\
&+& \frac{e(p+p')^\mu}{q\cdot(p+p')} 
[f^{-2}({p'}^2) S^{-1}_N(p') - f^{-2}(p^2)S^{-1}_N(p)]
+  \Ga^\mu_{\rm{amm}}.
\eea
This equation, together with \Eqref{formfnn}, 
defines the $\ga NN$ vertex of  the model.

One needs to keep in mind that since the free  Lagrangian is modified by form factors,
the propagators take the form
\beq
S^{\rm mod}_N(p) = f^2(p^2) S_N(p),
\eeq 
where $S_N(p) = (\slap-m_N)^{-1}.$ 
Nucleon spinors should be modified accordingly, \ie multiplied by $f$.
From \Eqref{modvert} it is particularly easy to see that the modified
vertex and propagator obey the same WT identity as the unmodified ones.
Thus, we have included the cutoff functions in  a way consistent with gauge
invariance.

Considering the pion case in the same fashion, and using the 
monopole form of $f(k^2)$, we find
\beq
 \Ga_{\ga \pi\pi}^{\mu,\rm{mod}}(k',k) = e (k+k')^\mu\left[ 
f^{-1}({k'}^2) f^{-1}(k^2) +f^{-1}({k'}^2) 
\frac{{k'}^2-m_\pi^2}{\La_\pi^2-m_\pi^2}
+ f^{-1}(k^2)\frac{k^2-m_\pi^2}{\La_\pi^2-m_\pi^2}\right].
\eeq
Note that the KR term is not modified, 
since we do not introduce any form factors
in the $\pi NN$ interaction Lagrangian, but rather have them in the propagators.

Since the modified
propagators and vertices obey the standard WT identities,
the proof of current conservation for the model with form factors
is exactly the same as before.

\section{Renormalization of the pole terms}

One of the effects of the \pn final-state interaction is to renormalize the $s$-channel contributions
of the photoproduction potential $V_{\pi \ga}$. 
Let us recall that the \pn amplitude $T_{\pi \pi}$ amplitude can symbolically
be presented as
\bea
T &=& \bGa^\dagger\, {\bf S}\, \bGa + T_u, \no\\
T_u &=& V_u+V_u G T_u,\no\\
\bGa &=&Z_1( \Ga+ \Ga G T_u),\\
{\bf S}^{-1}& =& S^{-1} - Z_1 \Ga G \bGa + Z_2(m-m_0) + (Z_2-1) S^{-1} \no\\
&=& Z_2 S^{-1}_0 - Z_1 \Ga G \bGa, \no
\eea
where $S^{-1}_0$ is the inverse bare propagator, e.g., for the nucleon it
is given by $\slap-m_0$.

The photoproduction potential $V_{\pi\ga}$ and the resulting amplitude 
$T_{\pi\ga}$
can also be separated into the ``pole'' and
``nonpole'' parts. In order for $T_{\pi\ga}$
to have the same dressed baryon exchanges  as in the $\pi\pi$ amplitude, one
ought to  use the bare parameters in the pole terms of the $V_{\pi\ga}$ potential, \ie
\beq
V_{\pi\ga} = \mbox{$\frac{Z_2}{Z_1}$}\Ga S_0 \Ga_\ga + V_{\pi\ga,u}\,,
\eeq
where $\Ga_\ga$ is the electromagnetic vertex.
Indeed, one then has
\bea
T_{\pi\ga}&=& (1+\bGa^\dagger\, {\bf S}\, \bGa\,G + 
T_u\,G) (\mbox{$\frac{Z_2}{Z_1}$}\,\Ga\, S_0\, \Ga_\ga + 
V_{\pi\ga,u})\no \\
&=& \bGa\, {\bf S}\, \Ga_\ga + T_u G V_{\pi\ga,u}.
\eea
We thus construct the nucleon- and $\De$-pole contributions by using the bare parameters
known from the $\pi N$ model, see Table~VII of Ref.~\cite{PaT00}.

\section{Results and discussion}

In \Figref{photo1} and \Figref{photo2} 
we present the model predictions for the
pion photoproduction multipoles, in units of $10^{-3}/m_{\pi^+}$.   
The dashed curves represent the Born amplitude without the sideways
form factors, while the dash-dotted curves -- with
the sideways form factors. The dotted curves show the
tree-level Born + $\rho, \w$ calculation with all the form factors
intact. The solid curves (Red -- real part, Blue -- imaginary part) 
represent the full calculation defined 
by \Eqref{trunc} (see also \Figref{gipif}) with Born + $\rho, \w, \De$
photoproduction potential and the complete $\pi N$ final state interaction from Ref.~\cite{PaT00}. 
The results are compared with the data from the partial-wave solutions of
Berends and Donnachie \cite{Ber75} and  SAID~\cite{SAID}.

The electromagnetic coupling parameters used 
in the calculation are given in Table~\tabref{gparams}, with $m_\w=0.783$
GeV, $\La_\w=\La_\rho$.
Only the $\De$-isobar electromagnetic couplings $g_M$ and $g_E$ 
were adjusted to for the best description
of the resonant multipoles: $M_{1+}^{(3/2)}$ and $E_{1+}^{(3/2)}$. In the figures 
we have plotted the results for the central values of these paramaters, given in bold in
Table~\tabref{gparams}. The other multipoles are very weakly
sensitive to the $\De$ isobar contribution (recall that the spin-1/2
backgrounds are absent in our model because of the specific form of the $\ga N\De$ vertex).
Other parameters have been taken from the literature~\cite{Kmat,Lag81}.

\begin{table}[h]
\begin{center}
\begin{tabular}{l}\hline \hline
$\kappa_V = { 3.71}, \kappa_S=\kappa_\w=-{0.12}$  \\
$g_{\rho NN}={ 2.66},\, g_{\w NN}={ 9.0},\,
\,g_{\ga\pi\w} = 3 g_{\ga\pi\rho} = { 0.313} $\\
$g_M = {\bf 2.8} \pm 0.2,\, g_E={\bf 1.5} \pm 0.5$ \\
\hline\hline
\end{tabular}
\end{center}
\caption{The electromagnetic coupling constants. The values given in bold were varied for a best fit.}
\tablab{gparams}
\end{table}

From the figures we see that the full model calculation
for most of the pion-photoproduction multipoles are in 
a very good agreement with the partial-wave analyses in this energy
region. The only problematic  $_pM_{1-}^{(1/2)}$ and $_nM_{1-}^{(1/2)}$
multipoles can possibly be corrected by including an explicit
Roper-resonance exchange in the photoproduction potential. It is expected to correct
not only the resonance but also lower energy region because of the N-Roper mixing
and related renormalization issues, cf.~\cite{PaT00} for details.

The difference
between the solid and dotted curves, in the non-resonant multipoles, can serve as a good measure
of the effect of the final state interaction. One can see that this effect is not dramatically large.
However it does make a significant difference in some channels, as will be demonstrated below.

Let us consider the reaction close to the threshold, $s\simeq (m_N+m_\pi)^2$.
The electric dipole amplitudes, $E_{0+}^{(I)}$, are of primary 
interest in this regime, all the other multipoles are tiny. There are predictions for
$E_{0+}$ from the low-energy theorems (LET) \cite{LET,BGKM}
and chiral perturbation theory (ChPT)~\cite{BKM96}. 
The result of the ``old'' LETs   \cite{LET} are given simply by the Born-graph contribution expanded in 
powers of $\mu=m_\pi/m_N$:
\begin{subequations}
\bea
\eqlab{oldlet}
E_{0+} (\pi^+ n) &= & \frac{e\, g_{\pi NN} }{8\pi m_N} \,\sqrt{2} \, (1-\thalf \mu) + O(\mu^2) \\
E_{0+} (\pi^- p) &= & \frac{e\, g_{\pi NN} }{8\pi m_N} \,\sqrt{2} \, (-1+\half \mu) + O(\mu^2)\\
E_{0+} (\pi^0 p) &= & -\frac{e\, g_{\pi NN} }{8\pi m_N} \, \mu [1-\half \mu (3+\kappa_p)] + O(\mu^3)\\
E_{0+} (\pi^0 n) &= & -\frac{e\, g_{\pi NN} }{8\pi m_N} \,\half \mu^2 \kappa_n + O(\mu^3)\,.
\eea
\end{subequations}
Bernard {\em et.al.}~\cite{BGKM} discovered that  at $O(\mu^2)$ there is
an important chiral-loop correction to the LET  for the {\it neutral} pion-production channels:
\begin{subequations}
\bea
\eqlab{newlet}
E_{0+} (\pi^0 p) &= & E_{0+}^{LET} (\pi^0 p)
+ \frac{e\, g_{\pi NN} }{8\pi m_N} \, \left(\frac{m_\pi}{4f_\pi}\right)^2 \\
E_{0+} (\pi^0 n) &= & E_{0+}^{LET} (\pi^0 n) 
+\frac{e\, g_{\pi NN} }{8\pi m_N} \, \left(\frac{m_\pi}{4f_\pi}\right)^2 \,,
\eea
\end{subequations}
where $f_\pi \simeq 93$ MeV is the pion decay constant. This result is commonly referred to as the
``new LET''.
Of course at this order there are also loop corrections to the charged multipoles, however they appear
to be less significant numerically than for the neutral channels.

The numerical values of these predictions,
together with the predictions of our model and some experimental results are collected
in Table~\tabref{dipoles}. In all of the theory predictions we have used $g_{\pi NN}^2/4\pi = 13.8$,
the value inferred from our pion-nucleon analysis.

\begin{table}[t]
\begin{center}
\begin{tabular}{|l|r|r||r|r|c|}\hline \hline
Multipole & Born  & $V_{\ga\pi}(1+G T_{\pi\pi})$ 
& old LET & new LET& Experiment \\ \hline
$E_{0+}({\pi^+\rn})$ & 26.1 &  $26.3$ & $25.9 $ 
& 25.9 & $27.9\pm 0.5$ (Ref.\cite{Ada66}), $28.06\pm 0.27$ (Ref.\cite{SAL99})\\
$E_{0+}({\pi^-\rp})$ & 
$-29.9$ &  $-29.6$ & $-30.8 $ & $-30.8$& $-31.4\pm 1.3$ (Ref.\cite{Ada66}),
$-31.5\pm 0.8 $ (Ref.\cite{Kov97})\\
$E_{0+}({\pi^0\rp})$ 
& $-2.4$ &  $-1.4$ & $-2.3$ & 1.0 
 & $-1.31\pm 0.08$ (Ref.\cite{Fuc96}), $-1.32\pm 0.05\pm 0.06$ (Ref.~\cite{Ber96}) \\
$E_{0+}({\pi^0\rn})$ & $0.4$ &  $1.0 $ & 0.5 & 3.8 & $\simeq 1.6 $  (Ref.~\cite{Ber97}) \\
\hline\hline
\end{tabular}
\caption{Predictions and experimental data for
the threshold electric dipole multipoles
for various reaction channels.}
\tablab{dipoles}
\end{center}
\end{table}

In Table~\tabref{dipoles}, the second column represent the value of the Born amplitude in our
model, while the third column corresponds to the full model calculations. It is reassuring that
without need to fit any parameters we obtained a reasonable agreement with experiment in all the
channels. It is also good to see that the effect of the FSI is small for the charged pion-photoproduction
and significant for the $\pi^0$ channels in analogy with the chiral loop effect of the new LET.
Thus, our results at threshold are in at least  qualitative agreement with ChPT.
They also are in reasonable quantitative agreement with experiment, and for the $\pi^0$ production
are even in better agreement than the ``new LET'' result. 
Although, it should be noted that in a more complete
calculation, including higher order effects and counter-terms, ChPT is in better agreement with
experiment than our simple model.  One can of course try to improve the model
by including higher-order contact
terms in the photoproduction potential. We however have not done that.
Our main aim is to apply the model in the resonance region
where ChPT is not applicable (yet). 

In particular, in the $\De$-resonance region we have been able to extract the coupling constants
of the $\ga N\rightarrow \De$ transition. 
A quantity of interest here
is the ratio of the electric   ($E2$) and magnetic ($M1$) $\ga N\rightarrow \De$
transition strength: $R_{EM}=E2/M1$. The physical significance of these value
is attributed to the deformation of the nucleon, see e.g.,~\cite{KMO84,Wal}.
For instance, in a naive quark model where the nucleon consists of
 three constituent quarks in the sphere-shape $S$ state -- the $E2/M1$ ratio vanishes.

In terms of the $\ga N\De$-vertex parameters
in our model the E2/M1 ratio is defined as (cf. Appendix A of Ref.~\cite{PP03}): 
 \beq
\eqlab{def1}
R_{EM}=  \frac{g_E}{2\,g_M\,\frac{m_{\Delta}+m_N}{m_{\Delta}-m_N}-g_E}  \, \times 100\%.
\eeq
Using the ``bare'' values of $g_M$ and $g_E$ in Table~\tabref{gparams}, we estimate this ratio
to be 
\beq
R_{EM}^{\rm (bare)} = (3.8 \pm 1.6)\,\%.
\eeq
We should immediately note that this value only seems to be inconsistent with PDG value~\cite{PDG}:
$R_{EM}=(-2.5\pm 0.5)\%$, the reason being that the PDG analyses define this ratio
as the ratio of corresponding resonant multipoles:
\beq
\eqlab{multdef}
R_{EM}^{\rm (multipoles)} = \frac{\im\, E_{1+}^{(3/2)}}{\im\, M_{1+}^{(3/2)}} \, \times 100\%.
\eeq
In our model we obtain $\im \,E_{1+}^{(3/2)} = -1.0 \pm 0.2$, and $\im \,M_{1+}^{(3/2)} = 38.5 \pm 1.5 $
(in units of $10^{-3}/m_{\pi}$)
at the $\De$ resonance position ({\it i.e.}, where Re$\,E_{1+}^{(3/2)} = 0=$ Re$\,M_{1+}^{(3/2)}$).
Therefore, we have
\beq
R_{EM}^{\rm (multipoles)} =(-2.6\pm 0.6 ) \,\%
\eeq
which is consistent with the PDG value.

The definition \eref{multdef},  however, is equivalent to \Eqref{def1} only assuming the 
that the on-mass-shell renormalized values of $g_M$ and $g_E$ are used in \Eqref{def1}.

Our result that the ``bare'' 
E2/M1 ratio is, in fact, small and {\it positive} is in agreement with other dynamical
models~\cite{DMT}, which allows us to believe that the model-dependence in the
extraction of this quantity in a dynamical modeling is rather mild and should be pursued further.

\section{Conclusion}

We have extended the dynamical modeling of the pion-nucleon system
in the first resonance region~\cite{PaT98,PaT00}
to the process of pion photoproduction on the nucleon.
Such extension is  indispensable in testing
the $\pi N$ dynamics beyond the elastic $\pi N$ scattering.

The presented numerical results are obtained
in the model which satisfies unitarity in the $\pi N\otimes \ga N$
channel space to the leading order in the electromagnetic coupling, 
and hence  Watson's theorem is exactly fulfilled.
We find that the model description of
the pion-photoproduction multipoles is in 
overall agreement with the partial-wave analyses in the region
from the threshold up to 650 MeV photon lab energy. 
We have therefore developed a realistic hadron-exchange model describing
the the low and intermediate energy 
pion scattering and photoproduction on the nucleon in a unitary fashion.
The model treats the quantum effects due to pion-nucleon loops in 
a Lorentz-covariant framework. It can be
extended to higher energies by including more reaction channels.
Furthermore, it is fully compatible
and complementary to the relativistic
meson-exchange models for the few-nucleon system, and hence
can naturally be embedded in these models to describe
more complicated processes.

 The results for
the threshold electric dipoles of the charged pion photoproduction
are very close to the low-energy theorem (LET) prediction and in a reasonable
agreement with experiment. In contrast, the electric dipole for the
neutral pion photoproduction off the proton, receives a sizable correction
due to the final state interaction and which improves the agreement with experimental as compared
to  LETs. This correction is found to be in a qualitative agreement with the large chiral-loop
correction to LET known from chiral perturbation theory (ChPT). 

The two parameters of the $\ga N\De$ vertex, which essentially 
are the only free parameters of the model, were fitted to $E_{1+}$ and $M_{1+}$
multipoles from the SAID solution. In future we plan to extract these parameters
directly from experimental data. At present,
the E2/M1 ratio obtained in the model is equal to $3.8\pm 1.6$ \% for the ``bare'' value
and to $-2.6\pm 0.6$ for the physical value. This is consistent with other
 analyses based on dynamical models.  The precise value of this ratio is model-dependent
as it is sensitive on the details of the $\pi N$ final state interaction.
The only possibility to extract this value is a model-independent way would be by using
ChPT with explicit $\De$ degrees of freedom. It would be extremely useful to carry out such
analysis.

\begin{acknowledgments}
We acknowledge valuable discussions with I.\ R.\ Afnan,  J.\ H.\ Koch,
A.\ Lahiff, and M.\ Vanderhaeghen at various stages of this work.
This work was supported by the U. S.
Department of Energy, under grants
DE-FG05-88ER40435, DE-FG02-04ER41302,  DE-FG02-93ER-40762, and DOE contract DE-AC05-84ER-40150 under
which the Southeastern Universities Research Association (SURA)
operates the Thomas Jefferson National Accelerator Facility.
\end{acknowledgments}

\appendix
\section{Lorentz, multipole, and isospin decomposition of the photoproduction amplitude}
The general Lorentz structure of
the fully off-shell $\ga\pi$ amplitude can be written as
\bea
\eqlab{offpiga}
T_{\la',s}^{\rho'\rho}(p',k';p,q) &=& \bar{u}_{\la'}^{\rho'} (\vp')\,
 (1, \slap') \left[ \bmat A_{11} &  A_{12}\\ 
  A_{21}  & A_{22}  \emat
+ \sla{P} \bmat B_{11} &  B_{12}\\ 
  B_{21}  & B_{22}   \emat \right. \no\\
&+& \left. \sleps \bmat C_{11} &  C_{12}\\ 
  C_{21}  & C_{22}  \emat
+ \sla{P}\,\sleps \bmat D_{11} &  D_{12}\\ 
  D_{21}  & D_{22}   \emat
\right] \barr 1\\ 
  \slap \earr \,u_{\la}^{\rho}(\vp), 
\eea
where $A,\,B\,,\,C\,, D\,$ are scalar functions, $\veps_\tau$ ($\tau=0,\pm 1$) stands for
the photon polarization vector,
index $s=\la-\tau$ denotes the helicity of the $\ga N$ state, $P$ is the total 4-momentum.

The parity-conserving $\ga\pi$ amplitudes are
the  transition amplitudes from the
$\ga N$ partial-wave state
\beq
\eqlab{pigastate}
\left. |J,r,\rho\right> = 
\frac{\left. |J,\rho,s\right> - r \rho
\left. |J,\rho,-s \right>}{\sqrt{2}}
\eeq
to the \pn partial-wave state: 
\beq
\eqlab{parstate}
\left. |J,r,\rho\right> = 
\frac{\left. |J,\rho,\la\right> + r\rho 
\left. |J,\rho,-\la \right>}{\sqrt{2}} .
\eeq
In terms of the partial-wave helicity amplitudes $M_{\la's}^{\rho'\rho}$, 
these amplitudes are given by
\beq
T^{\rho'\rho}_r = T_{\la's}^{\rho'\rho} - r T_{\la'\,-s}^{\rho'\rho}.
\eeq
For real photons $s$ takes the values: $-\thalf,-\half,\half,\thalf$.
Thus, for each parity $r$ and the $\rho$-spin values, we find two
independent amplitudes, e.g.,
\bea
&&{\mathscr A}_r^{\rho'\rho}=T_{\half \half}^{\rho'\rho} 
- r T_{\half\,-\half}^{\rho'\rho} ,\no\\
&&{\mathscr B}_r^{\rho'\rho}=T_{\half \thalf}^{\rho'\rho} 
- r T_{\half\,-\thalf}^{\rho'\rho} .
\eea

The {\it multipole amplitudes} are related to the parity
conserving amplitudes in the following way,
\begin{subequations}
\bea
E_{l+}&=& \frac{\sqrt{2}}{4(l+1)}\left[ 
\MA_+ + \sqrt{l/(l+2)}\,\MB_+\right] , \\
M_{l-}&=& \frac{\sqrt{2}}{4l}\left[ 
-\MA_- + \sqrt{(l-1)/(l+1)}\,\MB_-\right] , \\
E_{l-}&=& \frac{\sqrt{2}}{4l}
\left[\MA_- + \sqrt{(l+1)/(l-1)}\,\MB_- \right],\,\, J>1/2, \\
M_{l+}&=& \frac{\sqrt{2}}{4(l+1)}
\left[\MA_+ - \sqrt{(l+2)/l}\,\MB_+ \right],\,\, J>1/2, 
\eea
\end{subequations}
where $E$ or $M$ denotes whether the transition
is of electric or magnetic type. Index $l\pm$ stands for the value of the \pn state
orbital momentum, $l=J-\half r$, and the value of parity $r$.
 
Considering the isospin structure,
\beq
T={\pi'}^a \chi_N'  A_a \chi_N\,,
\eeq
the following three decompositions are usually made,
\bea
A_a &=& \de_{a3}\, A^{(+)} + \tau_a\, A^{(0)} +
i\veps_{a3b}\tau_b\, A^{(-)}  \no\\
&=&  \third \tau_a\tau_3 \, A^{(1/2)} + \tau_a\, A^{(0)}
+(\de_{a3}- \third \tau_a\tau_3)\, A^{(3/2)} \\
&=&  \half\tau_a(1+\tau_3) \, _pA^{(1/2)} + \half\tau_a(1-\tau_3) 
\, _nA^{(1/2)} + (\de_{a3}- \third \tau_a\tau_3)\, A^{(3/2)}. \no 
\eea
The relation amongst them is given by
\begin{subequations}
\bea
&&A^{(3/2)}=A^{(+)}-A^{(-)},\,\,\,\,
A^{(1/2)}=A^{(+)}+2A^{(-)}, \\
&& _pA^{(1/2)}=A^{(0)}+\third A^{(1/2)},\,\,\,\,
_nA^{(1/2)}=A^{(0)}-\third A^{(1/2)} . 
\eea
\end{subequations}
It is also possible to relate these to the 
amplitudes of specific reactions:
\begin{subequations}
\bea
&& A(\ga\rp\rightarrow \pi^0\rp)= A^{(+)} + A^{(0)}
=\mbox{$\frac{2}{3}$}A^{(3/2)} + {}_pA^{(1/2)}, \\
&& A(\ga\rn\rightarrow \pi^0\rn)= A^{(+)} - A^{(0)}
=\mbox{$\frac{2}{3}$}A^{(3/2)} - {}_nA^{(1/2)}, \\
&& A(\ga\rp\rightarrow \pi^+\rn)= \sqrt{2}(A^{(0)} + A^{(-)})
=\sqrt{2}(-\third A^{(3/2)} + {}_pA^{(1/2)}), \\
&& A(\ga\rn\rightarrow \pi^-\rp)= \sqrt{2}(A^{(0)} - A^{(-)})
=\sqrt{2}(\third A^{(3/2)} + {}_nA^{(1/2)}). 
\eea
\end{subequations}

\newpage

\begin{figure}[tb]
\centerline{  \epsfxsize=11 cm \epsffile{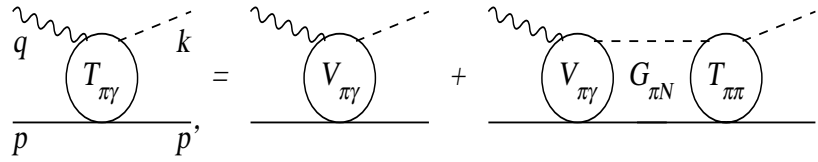}}
\caption{The unitary model for the photoproduction amplitude.}
\figlab{gipif}
\end{figure}
\bigskip

\begin{figure}[bt]
\centerline{  \epsfxsize=14 cm \epsffile{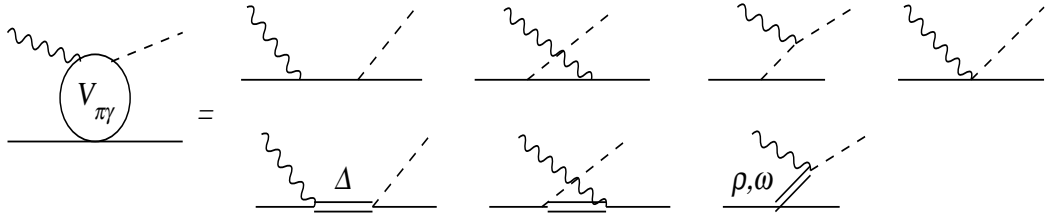}}
\caption{The tree-level photoproduction potential.}
\figlab{gapipotf}
\end{figure}

\epsfxsize=15cm
\begin{figure}[p]
\begin{center}
\centerline{\epsffile{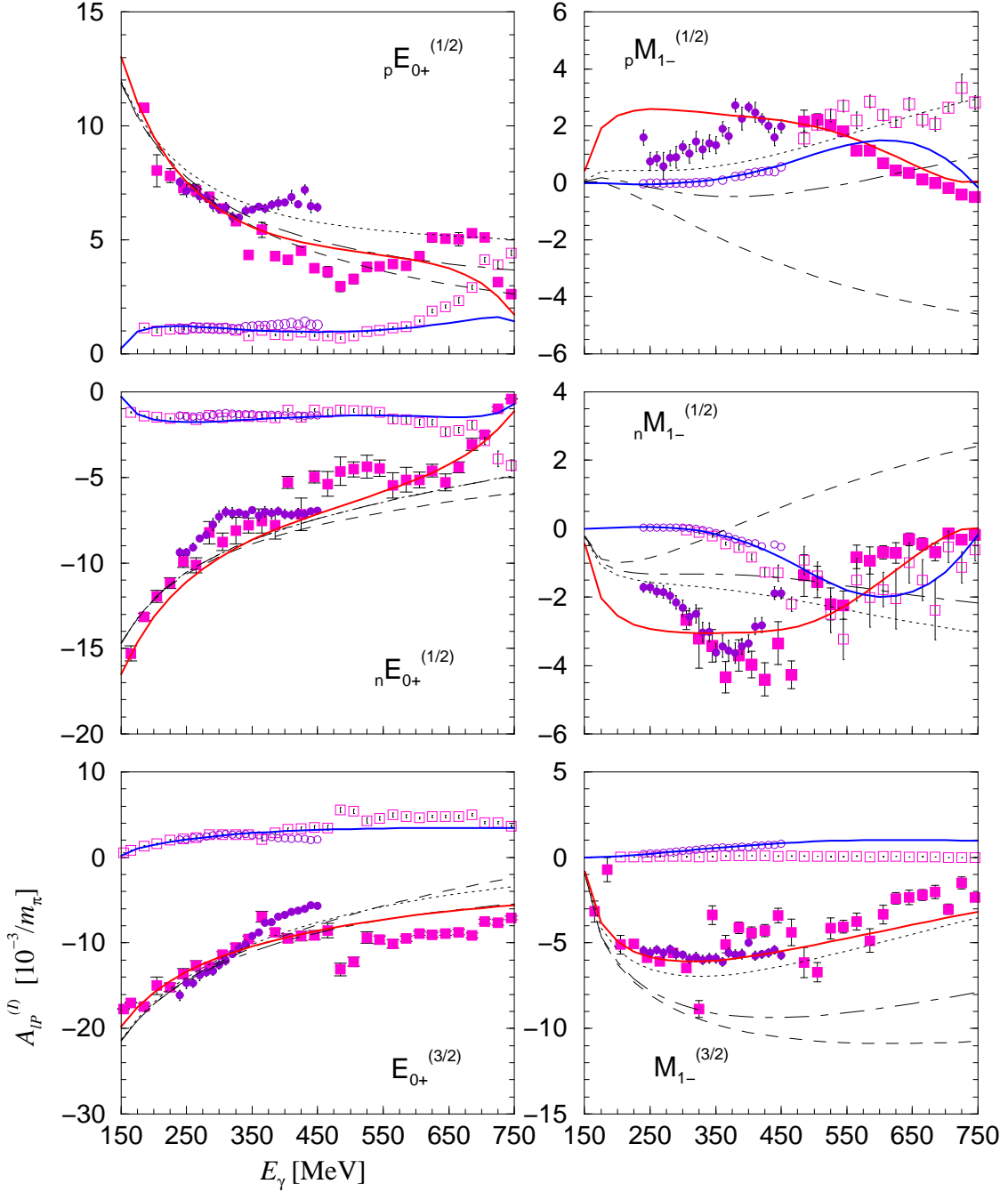}}
\end{center}
\caption{ (Color online)
The description of $J=1/2$ pion-photoproduction multipoles.
The dashed curves represent the Born amplitude {\it without} the sideways
form factors. The dash-dotted curves represent the Born amplitude 
{\it with} the sideways form factors. The dotted curves show the
tree-level Born + $\rho, \w$ calculation (with the form factors
intact). The solid curves are the full calculation including the
final state interaction (Re A -- red solid, Im A -- blue solid).
The results are compared to the partial-wave analyses: BD75 \cite{Ber75}
(Re A -- filled violet circles, Im A -- open violet circles), and SAID SM95 solution~\cite{SAID} (Re A -- filled 
purple squares, Im A -- open purple squares).} 
\figlab{photo1}
\end{figure}
\epsfxsize=15cm
\begin{figure}[p]
\begin{center}
\centerline{\epsffile{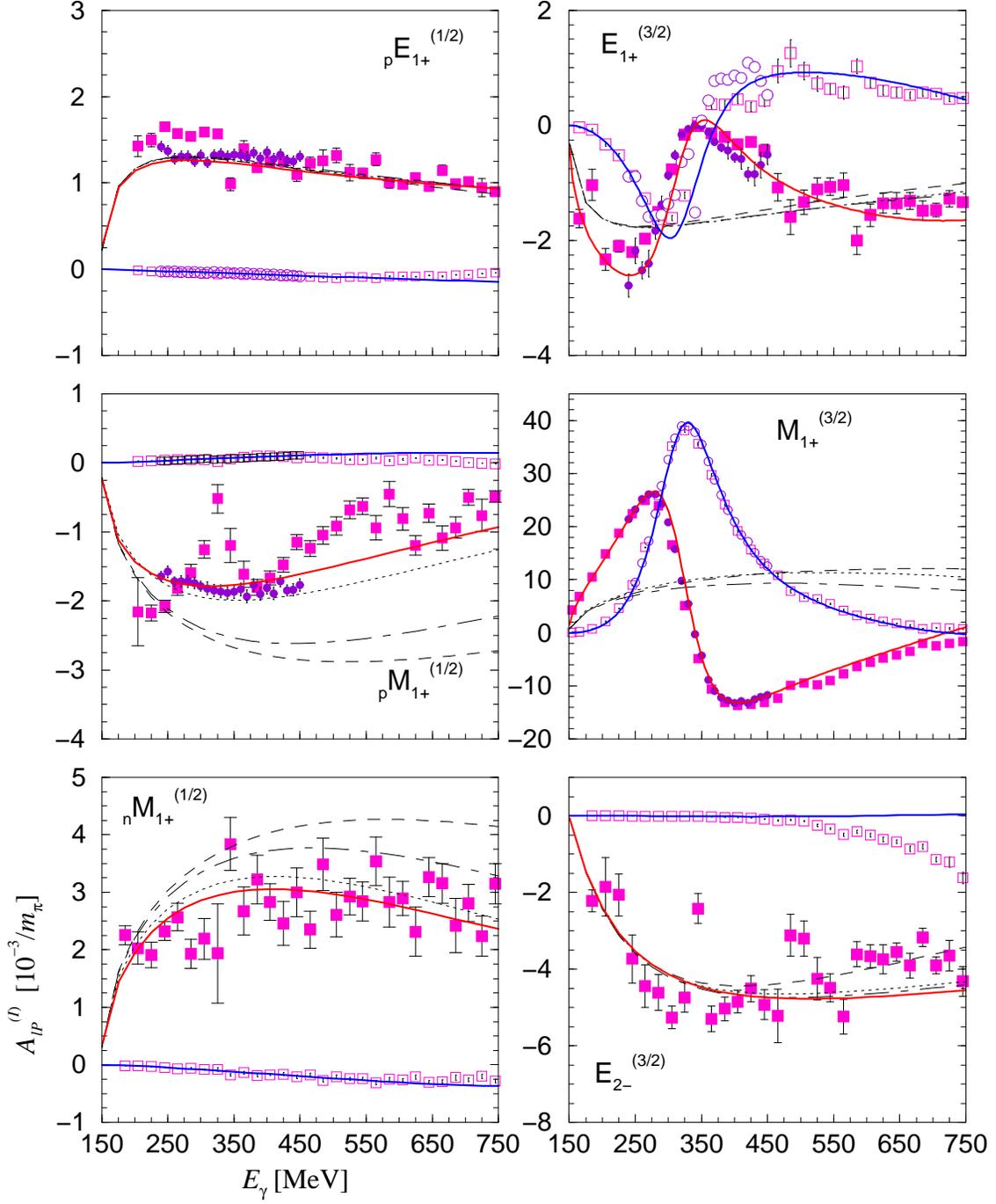}}
\end{center}
\caption{  (Color online) The description of some of the 
$J=3/2$ pion-photoproduction multipoles.
The legend is the same as in the previous figure.} 
\figlab{photo2}
\end{figure}
\end{document}